\documentclass[review]{elsarticle}
\usepackage{lineno,hyperref}
\usepackage{epsfig}
\usepackage{amsmath,amssymb,amsfonts}
\usepackage{mathrsfs}
\usepackage{bbm}
\usepackage{slashed}
\usepackage{graphicx}
\usepackage{verbatim}

\usepackage[T1]{fontenc}
\def\eq#1{(\ref{#1})}

\def\ba{\arraycolsep .1em \begin{eqnarray}}
\def\ea{\end{eqnarray}}

\newcommand{\be}{\begin{equation}}
\newcommand{\ee}{\end{equation}}
\newcommand{\bea}{\begin{eqnarray}}
\newcommand{\eea}{\end{eqnarray}}
\newcommand{\bi}{\begin{itemize}}
\newcommand{\ei}{\end{itemize}}

\newcommand{\p}{\partial}

\newcommand{\gb}{\bar{g}}

\newcommand{\cR}{\mathcal{R}}

\newcommand{\Ls}{M}

\begin{document}

\begin{frontmatter}

\title{Asymptotically safe cosmology -- a status report\footnote{Invited review for the special issue ``Testing quantum gravity with cosmology'' to appear in 
		Compte Rendus Physique A.}}

%% Group authors per affiliation:
\author{Alfio Bonanno}
\address{INAF, Osservatorio Astrofisico di Catania, via S. Sofia 78, I-95123 Catania, Italy \\
	INFN, Sezione di Catania, via S. Sofia 64, I-95123, Catania, Italy}
%\fntext[myfootnote]{Since 1880.}

\author{Frank Saueressig\footnote{Corresponding author.}}
\address{Institute for Mathematics, Astrophysics and Particle Physics (IMAPP), \\
Radboud University Nijmegen, Heyendaalseweg 135, 6525 AJ Nijmegen, The Netherlands}
%\fntext[myfootnote]{Since 1880.}

\begin{abstract}
Asymptotic Safety, based on a non-Gaussian fixed point of the gravitational renormalization group flow, provides an elegant mechanism for completing the gravitational force at sub-Planckian scales. At high energies the fixed point controls the scaling of couplings such that unphysical divergences are absent while the emergence of classical low-energy physics is linked to a crossover between two renormalization group fixed points. These features make Asymptotic Safety an attractive framework for cosmological model building. The resulting scenarios may naturally give rise to a quantum gravity driven inflationary phase in the very early universe and an almost scale-free fluctuation spectrum. Moreover, effective descriptions arising from an renormalization group improvement permit a direct comparison to cosmological observations as, e.g.\ \emph{Planck} data.
\end{abstract}


\end{frontmatter}

%\linenumbers
\newpage
%------------------------------------------------------------------------
\section{Asymptotic Safety: a brief introduction}
\label{sect1}
%------------------------------------------------------------------------
It is well-known that the quantization of general relativity based on the Einstein-Hilbert action results in a quantum field theory which is perturbatively non-renormalizable. This conclusion also holds if (non-supersymmetric) matter fields are added. The phenomenological success of general relativity then motivates to treat gravity as an effective field theory. This approach leads to a renormalizable theory of gravity in the sense that any quantum field theory becomes renormalizable if all possible counterterms compatible with its symmetries are included in the action \cite{Weinberg:2009bg}. While providing a consistent quantum theory for gravity, this construction falls short in terms of predictive power: while the effective field theory formulation works well at energy scales below the Planck scale where higher-derivative terms are suppressed by powers of the Planck mass, describing gravity at trans-Planckian scales requires fixing an infinite number of free coupling constants from experimental input.

In principle, Asymptotic Safety lives in the same space of theories as the corresponding effective field theory. It resolves the problem of ``predictivity'' encountered in effective field theory framework by imposing the extra condition that the quantum theory describing our world is located within the UV critical hypersurface of a suitable renormalization group (RG) fixed point. This condition implies that the high-energy behavior of the theory is controlled by the fixed point which renders all dimensionless coupling constants finite at high energy.  Fixing the trajectory uniquely then requires a number of experimental input parameters equal to the dimensionality of the hypersurface.

On this basis the crucial elements for Asymptotic Safety providing a valid theory for quantum gravity can be summarized as follows. Firstly, the existence of a suitable RG fixed point has to be shown. Secondly, the predictive power of the construction must be determined. Finally, it has to be shown that the UV critical hypersurface develops a regime where classical gravity constitutes a good approximation. Starting from the pioneering work \cite{Reuter:1996cp}, these points have been investigated in a vast variety of highly sophisticated computations, putting the scenario on firm grounds \cite{Niedermaier:2006wt,Codello:2008vh,Litim:2011cp,Percacci:2011fr,Reuter:2012id,Reuter:2012xf}. In particular the dimension of the UV critical hypersurface could be as low as three.

The prospect that Asymptotic Safety could be capable of describing gravitational force at all length scales makes the theory quite attractive for cosmological model building \cite{Bonanno:2001xi,Bonanno:2001hi,Bentivegna:2003rr,Reuter:2005kb,Bonanno:2007wg,Weinberg:2009wa,Bonanno:2009nj,Bonanno:2010mk,Koch:2010nn,Casadio:2010fw,Contillo:2010ju,Bonanno:2010bt,Frolov:2011ys,Hindmarsh:2011hx,Bonanno:2011yx,Ahn:2011qt,Cai:2011kd,Contillo:2011ag,Cai:2012qi,Bonanno:2012jy,Hindmarsh:2012rc,Fang:2012ca,Bonanno:2013dja,Copeland:2013vva,Kaya:2013bga,Becker:2014jua,Xianyu:2014eba,Xianyu:2014eba,Saltas:2015vsc,Nielsen:2015una,Bonanno:2015fga}. On the one hand some or all of the free parameters appearing in the Asymptotic Safety construction (including the value of the cosmological constant and Newton's constant complemented by a low number of higher-derivative couplings) may be determined from cosmological data. On the other hand, Asymptotic Safety provides a framework for developing effective cosmological models and 
 addressing questions related to a possible resolution of cosmological singularities. Typically, such investigations incorporate the effect of scale-dependent couplings through RG improvement techniques implemented either at the level of the equations of motion or the  effective (average) action. 
  While the resulting models are not based on the same level of rigor as the RG computations forming the core of the Asymptotic Safety program, they allow for the construction of interesting cosmological scenarios, e.g., in the framework of $f(R)$-type gravitational actions or dilaton-gravity theories. 

The rest of the work is then organized as follows. We briefly review the computation of gravitational RG flows and the central results in Sect.\ \ref{sect.2}, emphasizing the occurrence of a classical phase where general relativity is a good approximation. Cosmological models arising from RG improved equations of motion are discussed in  Sect.\ \ref{sect.3} while 
Sect.\ \ref{sect.4} summarizes results otained from (improved) effective actions.  We close with a brief summary and outlook in Sect.\ \ref{sect.5}. 

%------------------------------------------------------------------------
\section{Asymptotic Safety: fixed points and classical regime}
\label{sect.2}
%------------------------------------------------------------------------
Testing Asymptotic Safety at the conceptual level requires the 
ability to construct approximations of the gravitational RG flow beyond the realm of perturbation theory. A very powerful framework for carrying out such computations is  the 
functional renormalization group equation (FRGE) for the gravitational effective average action $\Gamma_k$ \cite{Reuter:1996cp}
\be\label{FRGE}
\p_k \Gamma_k[g, \gb] = \frac{1}{2} {\rm Tr}\left[ \left( \Gamma_k^{(2)} + \cR_k \right)^{-1} \p_k \cR_k \right] \, . 
\ee
The construction of the FRGE uses the background field formalism, splitting the metric $g_{\mu\nu}$ into a fixed background $\gb_{\mu\nu}$ and fluctuations $h_{\mu\nu}$. The Hessian $\Gamma_k^{(2)}$ is the second functional derivative of $\Gamma_k$ with respect to the fluctuation field at a fixed background and $\cR_k$ provides a scale-dependent mass term for fluctuations with momenta $p^2 \ll k^2$ with the RG scale $k$ constructed from the background metric. The interplay of $\cR_k$ in the numerator and denominator renders the trace both infrared and ultraviolet finite and ensures that the flow of $\Gamma_k$ is actually governed by fluctuations with momentum $p^2 \approx k^2$. In this sense, the flow equation realizes Wilson's idea of renormalization by integrating out ``short scale fluctuations'' with momenta $p^2 \ll k^2$ such that $\Gamma_k$ provides an effective description of physics for typical scales $k^2$. A priori one may then expect that resulting RG flow may actually depend strongly on the choice of background. As it was explicitly demonstrated in \cite{Benedetti:2010nr}, this is not the case, however: if flow is computed via early-time heat-kernel methods the background merely serves as a book-keeping device for disentangling the flow of different coupling constants.

The arguably simplest approximation of the gravitational RG flow is
obtained from projecting the FRGE onto the Einstein-Hilbert action approximating $\Gamma_k$ by
\be\label{EHans}
\Gamma_k = \frac{1}{16 \pi G_k} \int d^4x \sqrt{g} \left[-R + 2 \Lambda_k \right] + \mbox{gauge-fixing  and  ghost terms} \, . 
\ee
This ansatz comprises two scale-dependent coupling constants, Newton's constant $G_k$ and a cosmological constant $\Lambda_k$.
The scale-dependence of these couplings is conveniently expressed in
terms of their dimensionless counterparts
\be\label{dimless}
\lambda_k \equiv k^{-2} \, \Lambda_k \, , \quad g_k \equiv k^2 \, G_k \, ,
% \; \mbox{and} \; \; \, \eta_N \equiv (G_k)^{-1} \, k \p_k G_k \, . 
\ee
and captured by the beta functions
\be
k \p_k g_k = \beta_g(g_k, \lambda_k) \, , \qquad k \p_k \lambda_k = \beta_\lambda(g_k, \lambda_k) \, . 
\ee
Evaluating the beta functions \cite{Reuter:1996cp} for the Litim regulator \cite{Litim:2001up} gives
\be\label{beta2}
\begin{split}
\beta_\lambda = & \left(\eta_N -2 \right)\lambda + \frac{g}{12\pi}
\left[\frac{30}{1-2\lambda} - 24 - \frac{5}{1-2\lambda} \, \eta_N   \right] \\ 
\beta_g = & \left(2 + \eta_N \right) \, g \, ,
\end{split}
\ee 
with the anomalous dimension of Newton's constant $\eta_N \equiv (G_k)^{-1} \, k \p_k G_k$ being given by
\be
\eta_N = \frac{g \, B_1(\lambda)}{1 - g B_2(\lambda)}
\ee
where
\be
B_1(\lambda) = \tfrac{1}{3 \pi} \left[ \tfrac{5}{1-2\lambda} - \tfrac{9}{(1-2\lambda)^2} - 7 \right] \, , \quad 
B_2(\lambda) = - \tfrac{1}{12\pi} \left[\tfrac{5}{1-2\lambda} - \tfrac{6}{(1-2\lambda)^2} \right] \, . 
\ee

The beta functions \eqref{beta2} encode the scale-dependence of the dimensionless Newton's constant and cosmological constant. In particular, they contain the information on fixed points $g_*$ of the RG flow where, by definition of a fixed point, the beta functions vanish simultaneously, $\beta^a(g^a)|_{g^a = g_{*}^a} = 0$ constants. In the vicinity of a fixed point, the properties of the RG flow are captured by linearizing the beta functions around the fixed point. Defining the stability matrix ${\bf B}_{ab} \equiv  \p_{g^b}\beta_{g^a}|_{g = g_*}$  the linearized flow takes the form
\be
g^a(k) = g^a_* + \sum_{I} C^I \, V_I^a \, \left( \frac{k_0}{k} \right)^{\theta_I} \, .
\ee
Here the $V_I$ denote the right-eigenvectors of ${\bf B}$ with eigenvalues $-\theta_I$ such that $\sum_b {\bf B}_{ab} V^b_I = - \theta_I V ^a_I$, $k_0$ is a fixed reference scale and the $C^I$ are constants of integration. If Re$\theta_I > 0$ the flow along the eigendirection $V_I$ automatically approaches the fixed point $g^a_*$ as $k \rightarrow \infty$. In this case, the $C_I$ has a status of a free parameter. Analogously, eigendirections with Re$\theta_I$ < 0 are repelled from the fixed point as $k \rightarrow \infty$. The requirement that the fixed point controls the flow at high energy then demands that the corresponding integration constants $C_I$ must be set to zero. 
Compared to the effective field theory framework, Asymptotic Safety then potentially fixes an infinite number of free couplings, leading to a vast increase in predictive power.

The beta functions \eqref{beta2} give rise to two fixed points. Firstly, the Gaussian fixed point (GFP) is situated at $(g_*, \lambda_*) = (0,0)$. It corresponds to a free theory where the stability coefficients are determined by the mass-dimension of the coupling constant. Thus the GFP is a saddle point in the $g-\lambda$--plane: linearized solutions with $g > 0$ are repelled from this fixed point for $k\rightarrow \infty$. This feature reflects the perturbative non-renormalizability of the Einstein-Hilbert action in the Wilsonian language.

In addition, the flow possesses a non-Gaussian fixed point (NGFP) located at
\be\label{NGFP}
g_* = 0.707 \, , \qquad \lambda_* = 0.193 \, . 
\ee
From eq.\ \eqref{beta2} one sees that the anomalous dimension of Newton's constant at this fixed point is $\eta_N = -2$.
Its stability coefficients are given by
\be
\theta_{1,2} = 1.48 \pm 3.04i \, , 
\ee
such that RG flows in its vicinity actually spiral into the fixed point as $k \rightarrow \infty$. In the fixed point regime \eqref{NGFP} then entails that the dimensionful coupling constants scale according to
\be\label{FPscaling}
\lim_{k \rightarrow \infty} G_k = g_* \, k^{-2} \, , \qquad 
\lim_{k \rightarrow \infty} \Lambda_k = \lambda_* \, k^{2} \, . 
\ee
In particular the dimensionful Newton's constant vanishes as $k\rightarrow \infty$, entailing that the Asymptotic Safety mechanism renders gravity  anti-screening. 

\begin{figure}[t]
	\includegraphics[width=0.48\textwidth]{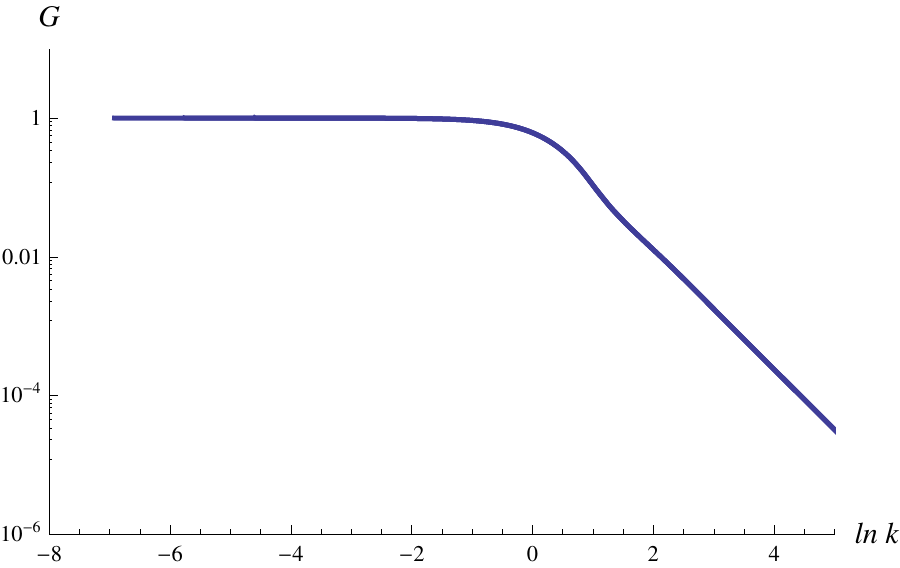} \; \;
	\includegraphics[width=0.48\textwidth]{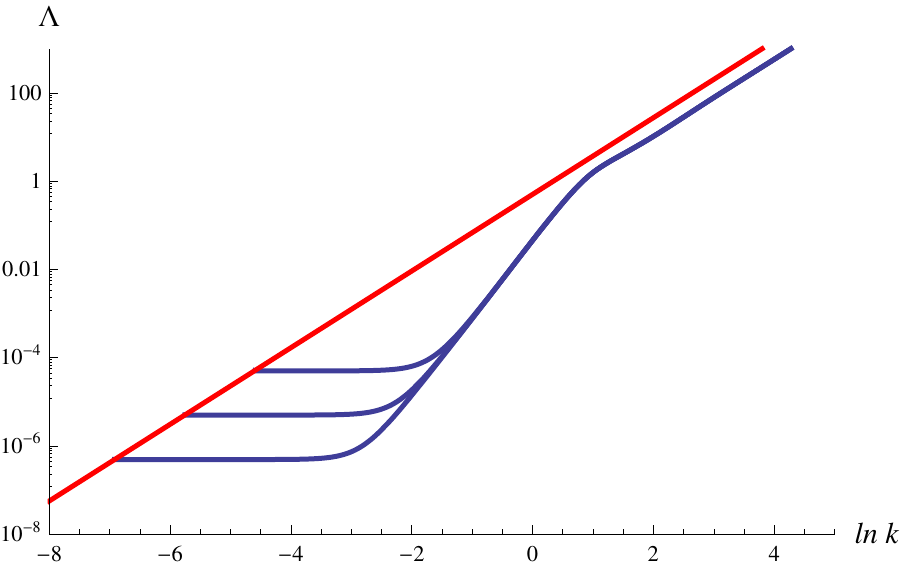}
	\caption{Scale-dependence of the Newton's constant (right) and cosmological constant (left) as a function of the RG scale $k$ for typical solutions giving rise to a positive cosmological constant. The red line indicates a singularity of the beta functions where $\eta_N$ diverges (adapted from \cite{Reuter:2001ag}).}
	\label{fig.1}
\end{figure}

At this stage it is instructive to construct the flow of $G_k$ and $\Lambda_k$
by integrating the beta functions \eqref{beta2} numerically. For solutions giving rise to a positive cosmological constant, typical examples are shown in Fig.\ \ref{fig.1}.
For $\ln k \gtrsim 2$ all solutions exhibit the fixed point scaling \eqref{FPscaling}. In the range $-2 < \ln k < 2$ the solutions undergo a crossover from the NGFP to the GFP. For $\ln k \lesssim -2$,  both $G_k$ and $\Lambda_k$ are (approximately) constant before terminating at finite $k_{\rm term}$ when $\lambda_{k_{\rm term}} \approx 1/2$. The value of $G_k$ and $\Lambda_k$ in this classical regime can be set to the experimentally measured values $G_0$ and $\Lambda_0$ by choosing suitable initial conditions \cite{Reuter:2004nx}. The small value of $\Lambda_0$ then ensures that the classical regime extends from the Planck scale up to cosmic scales. 

At this stage the following remarks are in order. The result \eqref{NGFP} actually constitutes the \emph{projection} of the NGFP underlying Asymptotic Safety to the space of interactions spanned by the Einstein-Hilbert action. It is unlikely that the full effective average action $\Gamma_*$ is of Einstein-Hilbert form. Starting from the Einstein-Hilbert result, the existence of a suitable NGFP has by now been established in a series of 
highly sophisticated approximations, reviewed, e.g., in \cite{Reuter:2012id}.
The inclusion of power-counting marginal four-derivative terms thereby indicate that there is (at least) one additional relevant direction which may be associated with the coupling constant associated with an $R^2$-interaction \cite{Lauscher:2002sq,Benedetti:2009rx}. Notably, the existence of a classical regime persists upon including higher-derivative terms comprising $R^2$-interactions or the Goroff-Sagnotti counterterm \cite{Rechenberger:2012pm,Gies:2016con}. The mechanism giving rise to this feature is universal: the classical regime  results from a crossover from the NGFP controlling the high-energy behavior to the GFP governing the low-energy physics. This crossover also works for realistic values of the cosmological constant \cite{Reuter:2004nx}.

%------------------------------------------------------------------------
\section{Quantum gravity effects and the initial singularity}
\label{sect.3}
%------------------------------------------------------------------------
In principle, astrophysical and cosmological applications of the asymptotic safety scenario are ubiquitous. Fig.\ \ref{fig.1} displays a significant scale-dependence of Newton's constant and the cosmological constant for trans-Planckian energy outside the classical regime. This raises the immediate question if this scale-dependence has an influence on cosmic singularities and if it leaves phenomenological imprints in cosmological signatures. These questions may be addressed using RG improvement techniques reviewed below.

%------------------------------------------------------------------------
\subsection{Incorporating quantum gravity via renormalization group improvements}
\label{sect.3a}
%------------------------------------------------------------------------
A key feature of the effective average action $\Gamma_k[g_{\mu\nu}]$ is that it provides an effective description of the physical system at scale $k$. 
Based on this property $\Gamma_k$ allows to derive effective field equations for the effective metric 
via 
\be
\label{EOM}
\frac{\delta \Gamma_k}{\delta g_{\mu\nu}(x)}[\langle g\rangle_k]=0  
\ee
where, quantities $\langle \cdot \rangle$ can be interpreted as averaged
over (Euclidean) spacetime volumes with a linear extension of order $k^{-1}$.
In the case of Einstein-Hilbert truncation \eqref{EHans}, one obtains
\be\label{effective}
R_{\mu\nu}[\langle g \rangle_k] - \frac{1}{2} R\,\langle g_{\mu\nu}\rangle_k = - \Lambda(k) \, \langle g_{\mu\nu}\rangle_k + 8\pi G(k) \, \langle T_{\mu\nu}\rangle_k \, ,
\ee
with $T_{\mu\nu}$ the standard energy momentum tensor describing the matter content of the system. In the RG improvement process the cutoff $k$ is then identified with a typical length scale of the system, $k \mapsto k(x^\mu)$. In the context of cosmology, there are several types of cutoff identifications   
\begin{subequations}\label{cutoffids}
	\begin{align}
& \mbox{Type I:} \quad	   &    k^2 & = \xi^2 \, t^{-2} \, ,  \label{typeI} \\
& \mbox{Type II:} \quad   &	k^2 &= \xi^2 \, H(t)^2 \, , \label{typeII}\\
& \mbox{Type III:} \quad  &	k^2 &= \xi^2 \,\sqrt{R_{\mu\nu\rho\sigma}R^{\mu\nu\rho\sigma}} \, ,  \label{typeIII}\\
& \mbox{Type IV:} \quad   &	k^2 & = \xi^2 \,T^2 \, .  \label{typeIV}
	\end{align}
\end{subequations}
Here $t$ denotes cosmic time (representing a proper distance), $H(t)$ the Hubble parameter, $\sqrt{R_{\mu\nu\rho\sigma}R^{\mu\nu\rho\sigma}} $ is representative for a quantitiy characterizing the curvature of spacetime, and $T \propto \rho^{1/4}$ is the temperature of the cosmic plasma. Moreover, $\xi$ is an a priori undetermined positive parameter of order one. Supplementing \eqref{effective} by a suitable equation of motion for the matter sector and substituting one of the cutoff identifications leads to a closed system of equations which allows to determine the averaged metrics $\langle g_{\mu\nu}\rangle_k$. If $\Lambda(k)$ and $G(k)$ are (approximately) scale-independent the dynamics entailed by \eqref{effective} reduces to the one of general relativity while the running of the couplings induces distinct modifications controlled by the beta functions of the theory. Fig.\ \ref{fig.1} then indicates that these corrections will set in when $k^2 \gtrsim G_0$ which is the natural scale for quantum gravity effects. 

An alternative to the RG improvement of the equations of motion described above can be borrowed from QED and QCD \cite{Migdal:1973si,Adler:1982js,Dittrich:1985yb} and carries out the improvement at the level of the effective action. Instead of calculating the effective action in terms of Schwinger's proper-time approach or perturbative calculations of Feynman diagrams, it turns out to be more convenient to 
obtain the low energy effective action by means of the stress-energy tensor and 
the leading-log model. 
In a similar way the RG approach to  gravity allows to construct improved actions by promoting $k$ to an external spacetime dependent field $k=k(x^\mu)$
 or identifying $k$ directly with the field strength. The later corresponds to  a cutoff identification of Type III applied to the effective average action. This procedure then leads to additional terms in the equations of motion which originate from $D_\mu G(k(x)) \not = 0$ (also see \cite{Reuter:2004nx,Koch:2010nn} for a more detailed discussion).

%------------------------------------------------------------------------
\subsection{Friedmann-Robertson-Walker cosmology}
%------------------------------------------------------------------------
A natural starting point for investigating potential signatures of Asymptotic Safety studies the RG improved equations of motion for homogeneous and isotropic flat Friedmann-Robertson-Walker cosmologies. In this case complete cosmic histories taking the scale-dependence of the couplings into account have been developed in a series of works \cite{Bonanno:2001xi,Reuter:2005kb,Bonanno:2007wg,Koch:2010nn}.\footnote{Also see \cite{Bonanno:2011yx,Kofinas:2016lcz} for related discussions.}
In this case the line element
\be
ds^2 = - dt^2 + a(t)^2 \left[ dx^2 + dy^2 + dz^2 \right] \, , 
\ee
is supplemented by a stress-energy tensor of a perfect fluid, $T_\mu{}^\nu = {\rm diag}[-\rho, p,p,p]$, satisfying the equation of state $p = w \rho$. Applying a cutoff identification $k \mapsto k(t)$ to \eqref{effective} leads to the RG improved Friedmann and continuity equation
\be\label{impFRW}
\begin{split}
	&	H^2 = \frac{8\pi}{3} G(t) \rho + \frac{1}{3} \Lambda(t) \, , \\ 
	&	\dot{\rho} + 3 H (\rho + p) =  - \frac{\dot{\Lambda} + 8 \pi \rho \, \dot{G}}{8 \pi G(t)} \, . 
\end{split}
\ee
The second equation arises from the Bianchi identity satisfied by Einstein's equations $D^\mu[\lambda(t) \, g_{\mu\nu} - 8 \pi G(t) \, T_{\mu\nu}]=0$. The extra term on its right-hand-side has the interpretation of an energy transfer between the gravitational degrees of freedom and matter. Introducing the critical density $\rho_{\rm crit} \equiv 3 H(t)^2/(8 \pi G(t))$ and defining the relative densities $\Omega_{\rm matter} = \rho/\rho_{\rm crit}$ and $\Omega_\Lambda = \rho_\Lambda/\rho_{\rm crit}$ the first equation is equivalent to $\Omega_{\rm matter} + \Omega_\Lambda = 1$.

We first focus on the very early part of the cosmological evolution where the scaling of $G$ and $\Lambda$ is given by \eqref{FPscaling}. Selecting the cutoff identification to be of Type II, eq.\ \eqref{typeII}, 
 the system \eqref{impFRW} has the analytic solution
\be\label{NGFPsol}
H(t) = \frac{\alpha}{t} \, , \qquad a(t) = A \, t^\alpha \, , \qquad \alpha = \left[\tfrac{3}{2} (1+w) (1-\Omega^*_\Lambda) \right]^{-1} \, , 
\ee
together with $\rho(t) = \widehat{\rho} \, t^{-4}$, $G(t) = \widehat{G} \, t^2$ and $\Lambda(t) = \widehat{\Lambda} \, t^{-2}$. The constants $\widehat{G},\widehat{\Lambda}$ and $\widehat{\rho}$ are determined in terms of the position of the NGFP and the free parameter $\xi$,
\be
\widehat{\rho} = \tfrac{3}{8\pi} \tfrac{\xi^2 \alpha^4}{g_*} \left(1-\tfrac{1}{3} \lambda_* \xi^2 \right) \, , \quad \widehat{G} = \frac{g_*}{\xi^2 \alpha^2} \, , \quad \widehat{\Lambda} = \lambda_* \, \xi^2 \, \alpha^2 \, , \quad \Omega_\Lambda^* = \tfrac{1}{3} \, \lambda_* \, \xi^2 \, , 
\ee
while $A$ is a positive constant. The vacuum energy density in the fixed point regime, $\Omega_\Lambda^*$ takes values in the interval $]0,1[$.\footnote{A priori the value of $\alpha$ depends on the parameter $\xi \simeq O(1)$ entering the renormalization group improvement scheme \eqref{cutoffids}. For radiation dominance $w=1/3$, and the fixed point \eqref{NGFP}, one typically has $1/2 < \alpha < 1$. In principle, the value of $\xi$ may be fixed by imposing, e.g., conservation of the classical stress-energy tensor \cite{Bonanno:2001xi}, but we will consider $\xi$ as a free parameter in the sequel.}
 The solutions \eqref{NGFPsol} possess no particle horizon if $\alpha \ge 1$ while for $\alpha < 1$ there is a horizon of radius $r_H = t/(1-\alpha)$. Moreover, they 
undergo power law inflation if $\alpha > 1$.  Assuming radiation dominance, $w = 1/3$, this requires $\Omega_\Lambda^* > 1/2$. For the NGFP \eqref{NGFP} this corresponds to $2.79 \le \xi \le 3.94$. Remarkably, the asymptotic behavior of the solution for $t \rightarrow 0$ is actually independent of the chosen improvement scheme: given the solution \eqref{NGFPsol} together with the curvature tensor evaluated in Tab.\ \ref{tab.1}, all choices entail $k \propto t^{-1} + \mbox{subleading}$, corroborating the robustness of the improvement procedure.

Realizing an inflationary phase in the fixed point regime by having $\Omega_\Lambda^* \ge 1/2$ is a rather attractive scenario: inflation driven by the quantum gravity effects ends automatically at the transition time $t_{\rm tr}$ when the RG flow enters into the classical regime. For $t > t_{\rm tr}$  the evolution is then given by a classical Friedmann-Robertson-Walker universe. The period of a NGFP-driven inflationary phase does not require any extra ingredients like an inflaton or a specific inflaton potential. 

Quite remarkably, the NGFP-driven inflation may leave imprints in the cosmic fluctuation spectrum. The transition time $t_{\rm tr}$ is determined by the scale $k$ where the underlying RG trajectory enters into the classical regime $G_0$. From Fig.\ \ref{fig.1} one sees that $k \simeq m_{\rm Pl}$ where the Planck mass is determined from the value of Newton's constant in the classical regime. Since $\xi = O(1)$ gives $H(t_{\rm tr}) \approx m_{\rm Pl}$. The relation $H(t) = \alpha/t$ then leads to the estimate
\be
t_{\rm tr} = \alpha \, t_{\rm Pl} \, . 
\ee
If $\Omega_\Lambda^*$ is very close to one, i.e., $\alpha \gg 1$ the cosmic time $t_{\rm tr}$ when the Hubble parameter is of order $m_{\rm Pl}$ can be much larger than the Planck time which is then located within the NGFP regime.

We now consider the evolution of a fluctuation with comoving length $\Delta x$. The corresponding physical length is $L(t) = a(t) \Delta x$. In the NGFP regime, $L(t)$ is related to the proper length at the transition time $t_{\rm tr}$ via $L(t) = (t/t_{\rm tr})^\alpha \, L(t_{\rm tr})$. The ratio of $L(t)$ and the Hubble radius $\ell_H(t)$ then evolves as 
\be\label{mag1}
\frac{L(t)}{\ell_H(t)} = \left(\frac{t}{t_{\rm tr}}\right)^{\alpha-1} \, \frac{L(t_{\rm tr})}{\ell_H(t_{\rm tr})} \, . 
\ee
For $\alpha > 1$ the proper length of the object grows faster than the Hubble radius. Fluctuations which are of sub-Hubble size at early times can then cross the horizon and become ``super-Hubble''-size at later times.

For definiteness, let us consider a fluctuation which, at the transition time $t_{\rm tr}$ is $e^N$ times larger than the Hubble radius. For this fluctuation, eq.\ \eqref{mag1} implies 
\be
\frac{L(t)}{\ell_{\rm H}(t)} = e^N \left( \frac{t}{t_{\rm tr}} \right)^{\alpha -1} \, . 
\ee
The time $t_N$ where this fluctuation crosses the Hubble horizon, $L(t_N) = \ell_H(t_N)$ is
\be\label{tNeq}
t_N = t_{\rm tr} \, \exp\left( - \frac{N}{\alpha-1}\right) \, . 
\ee
Thus even for moderate values of $\alpha$, NGFP-driven inflation easily magnifies fluctuations to a size where they are many orders of magnitude larger than the Hubble radius. Interestingly, the structures visible today may have crossed the Hubble horizon during the NGFP regime. Starting from the largest structures visible today and using the classical evolution to backtrace them in time to the point where $H = m_{\rm Pl}$ their size back then is given by $e^{60} \ell_{\rm Pl}$. Setting $N=60$ the time $t_{60}$ when these structures crossed the horizon can be estimated from eq.\ \eqref{tNeq}. For $\alpha = 25$,  $t_{60} = t_{\rm tr}/12.2 = 2.05 t_{\rm Pl}$. Thus $t_{60}$ is one order of magnitude smaller than $t_{\rm tr}$. In this setting structures observed today may have their origin in the quantum regime controlled by the NGFP.

The NGFP also offers a natural mechanism for generating a scale-free spectrum of primordial fluctuations \cite{Bonanno:2007wg}. Following the discussion \cite{Antoniadis:1996dj}, this can be seen as follows: owed to the anomalous dimension of the theory at the NGFP, $\eta_N = -2$, the effective graviton propagator (at background level) has a characteristic $1/p^4$ dependence. 
This implies a  logarithmic dependence for the two-point graviton correlator in the configuration space, $\langle h_{\mu\nu}(x)h_{\mu\nu}(y)\rangle \sim \ln(x-y)^2$.
As a consequence curvature fluctuations $\delta {\bf R}\propto \partial^2 h$ (where $\bf R$ stands for any component of Riemann or Ricci tensor)
must behave as:
$\langle \delta {\bf R}({\bf x},t) \delta {\bf R}({\bf y},t) \rangle \propto 1/|{\bf x}-{\bf y}|^4$.
If the fluctuations on the matter part $\delta \rho$ originate from the fluctuations of the geometry itself, the classical Einstein equations provide the relation $\delta \rho \propto \delta \bf R$. 
The correlation function $\xi({\bf x}) \equiv \langle \delta({\bf x}) \delta(0) \rangle$ of the density contrast of $\delta({\bf x}) \equiv \delta\rho({\bf x})/\rho$ must behave as \cite{Bonanno:2007wg}
\be
\label{corr}
\xi({\bf x})\propto \frac{1}{|{\bf x}|^4}
\ee
provided the physical distance $a(t)|{\bf x}|$ is smaller than the Planck length. Therefore, from the 3-dimensional Fourier
transform of \eq{corr} we immediately get $|\delta_k|^2\propto |{\bf k}|$ which results in a scale invariant power spectrum, with 
spectral index $n=1$. Clearly (small) deviations from $n=1$ are expected as, strictly speaking, the prediction of an exactly scale-free spectrum holds at the NGFP only. Since the NGFP is supposed to govern the dynamics of the theory before the onset of inflation this implies in particular that the power spectrum may acquire non-trivial corrections during the inflationary phase.

%------------------------------------------------------------------------
\subsection{BKL-Type singularities}
%------------------------------------------------------------------------
%
\begin{figure}[t]
	\includegraphics[width=0.48\textwidth]{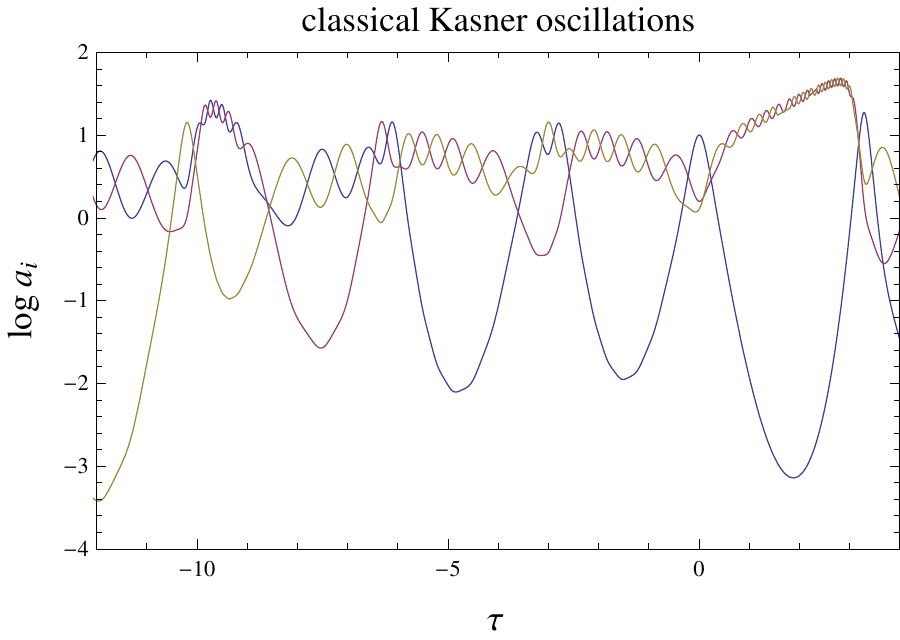} \; \;
	\includegraphics[width=0.48\textwidth]{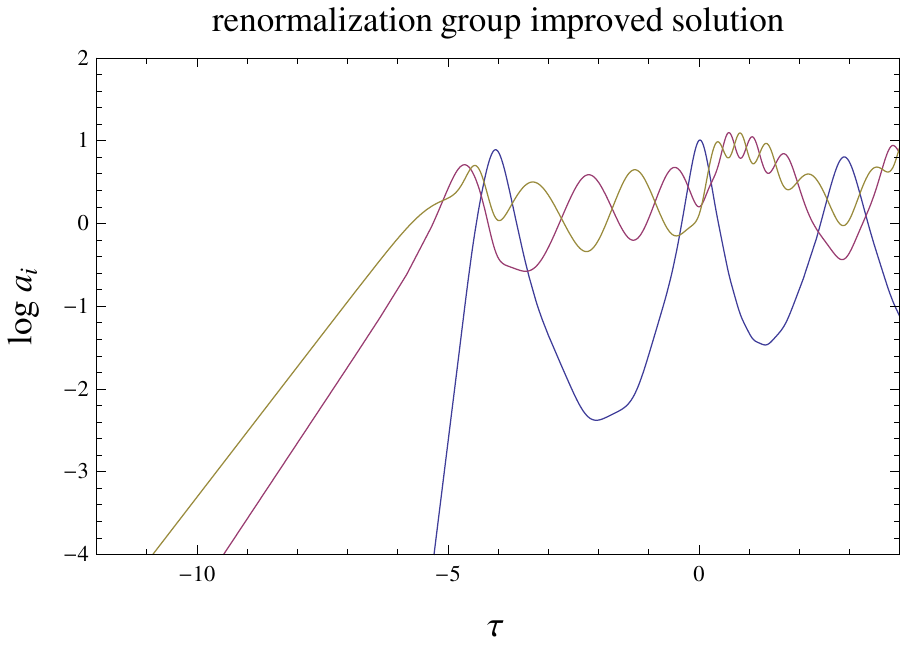}
	\caption{\label{fig.2} Dynamics of the scale factors $a_i$ for the classical (left) and RG improved (right) vacuum Bianchi IX model as a function of $d\tau \equiv a_1 a_2 a_3 \, dt$. The RG improvement leads to a decoupling of spatial points such that the system enters into a quiescence phase (adapted from \cite{D'Odorico:2015lhd}).}
\end{figure}
The RG improvement techniques employed in the case of a homogeneous and isotropic Friedmann-Robertson-Walker solution are readily extended
 anisotropic models \cite{D'Odorico:2015lhd}. This includes the class of vacuum Bianchi I and Bianchi IX models where the line element has the form
\be\label{BianchiIX}
ds^2 = -dt^2 + \left(a^2_1 \, l_a l_b + a^2_2 \, m_a m_b + a^2_3 \, n_a n_b \right)dx^a dx^b \, . 
\ee
Here $a_i(t)$ are scale factors depending on the cosmological time $t$ and the  three-vectors $l_a$, $m_a$, $n_a$ depend on the spatial coordinates and determine the directions scaling with the corresponding $a_i$. At the classical level, BKL \cite{Belinsky:1970ew,Belinski:1973zz,Belinsky:1982pk} discovered that the dynamics of the scale factors follows a complex oscillatory pattern between Kasner phases where the spatial derivatives of the three-vectors are negligible (see right panel of Fig.\ \ref{fig.2}).
In a Kasner phase the scale factors follow a power-law behavior $a_i(t) = t^{p_i}$ with the Kasner exponents $p_i$ satisfying
\be
\sum_{i=1}^3 \, p_i = 1 \, , \qquad \sum_{i=1}^3 \, (p_i)^2 = 1 \, . 
\ee
The solution of these equations may be parameterized in terms of a single real variable $u$, showing that classically one always has two positive and one negative Kasner exponent (the point $p_1 = 1$ and $p_2 = p_3 = 0$ constituting an exception). If the system approaches $t \rightarrow 0$, the scale-factor associated with the negative Kasner exponent becomes large, triggering a bounce into a new Kasner phase. Thus classically, the system undergoes an infinite number of Kasner bounces as it approaches the initial singularity at $t = 0$.
\begin{table}[t!]
	\begin{center}
		\begin{tabular}{|c|c|}
			\hline
			\hline
			\hspace{5mm}	Singularity 	\hspace{5mm} & 	\hspace{5mm} $R_{\mu\nu\rho\sigma}R^{\mu\nu\rho\sigma}$ 	\hspace{5mm}  \\ \hline
			FRW cosmology & $\frac{12 \alpha^2 (1-2\alpha+2\alpha^2)}{t^4}$ \\ 
			BKL singularity & $\frac{4\left[r + \lambda_* + (r+\lambda_*)^2 - 2 (p_1^3+p_2^3+p_3^3) - p_1^2p_2^2 - p_1^2p_3^2 - p_2^2 p_3^2\right]}{t^4}$ \\
			\hline \hline
		\end{tabular}
		\caption{\label{tab.1} Initial singularities for the RG improved Friedmann-Robertson-Walker (FRW) solution (top) and the vacuum Bianchi IX universe (bottom). The values $\alpha$ and $p_i$ determining the square of the Riemann tensor are given in eqs.\ \eqref{NGFPsol} and \eqref{psol}. Both models exhibit a point singularity at $t=0$.}
	\end{center}
\end{table}

We now include a scale-dependent cosmological constant and perform a Type I RG improvement identifying $k^2 = \xi^2 t^{-2}$. In the NGFP regime the improved vacuum equations of motion take the form
\be
R_{\mu\nu} - \tfrac{1}{2} g_{\mu\nu} R = - \lambda_* \, t^{-2} \, g_{\mu\nu} \, ,
\ee
where we absorbed $\xi^2$ into the parameter $\lambda_*$. Neglecting spatial gradients (corresponding to the Bianchi I case), the system again possesses Kasner-type scaling solutions where the Kasner exponents satisfy
\be\label{Kasnerimp}
\sum_{i=1}^3 \, p_i = r \, , \qquad \sum_{i=1}^3 \, (p_i)^2 = r + \lambda_* 
\ee
with $r \equiv (1+\sqrt{1+12\lambda_*})/2$. The one-parameter family of solutions of this system is conveniently parameterized by $u \in [0,1]$ and given by 
\be\label{psol}
\begin{split}
p_1 = & \tfrac{1}{3}\left(r-\sqrt{r}\right) - \frac{\sqrt{r} \, u}{1+u+u^2} \, , \\
p_2 = & \tfrac{1}{3}\left(r-\sqrt{r}\right) + \frac{\sqrt{r} \, u(1+u)}{1+u+u^2} \, , \\
p_3 = & \tfrac{1}{3}\left(r-\sqrt{r}\right) + \frac{\sqrt{r} \, (1+u)}{1+u+u^2} \, . \\
\end{split}
\ee
For $\lambda_* > 0$, these solutions possess the remarkable feature that there are intervals $u$ where all Kasner exponents are positive.

Returning to the Bianchi IX case with the spatial gradients turned on, a generic solution undergoes Kasner oscillations. When tracing the dynamics back in time, the crucial difference to the classical case occurs if a Kasner bounce reflects the system into the part of the solution space where all $p_i > 0$. In this phase there is then no scale factor which diverges as $t \rightarrow 0$. Consequently, the Kasner bounces stop and the system approaches a point-like singularity where $\lim_{t \rightarrow 0} a_i(t) = 0$, $i = 1,2,3$ (see the right panel of Fig.\ \ref{fig.2} for illustration). The RG improved
Bianchi IX model exhibits the same quiescent behavior found when the Bianchi IX universe is populated by stiff matter \cite{Andersson:2000cv}. Moreover, it  gives rise to the same type of point singularity as the one encountered in the homogeneous and isotropic case.\footnote{In the context of shape dynamics a similar setup has recently been studied in \cite{Koslowski:2016hds} where it was shown that the specific properties underlying shape dynamics allow the continuation of solutions through this singularity.}

%------------------------------------------------------------------------
\section{Inflationary models}
\label{sect.4}
%------------------------------------------------------------------------
A more detailed connection between Asymptotic Safety and cosmological data may be obtained through the construction of effective actions valid at the scale of inflation. For pure gravity the effective actions studied so far fall into the class of $f(R)$-type theories where the modifications are attributed to quantum gravity effects. Along a different line it is also possible to extend 
the pure gravity theory by including an additional scalar field and investigate imprints of Asymptotic Safety within the framework of dilaton-gravity theories. These two cases  will be discussed in subsections \ref{sect.4a} and \ref{sect.4b}, respectively.

%------------------------------------------------------------------------
\subsection{Effective $f(R)$-type gravity models}
\label{sect.4a}
%------------------------------------------------------------------------
According to the inflationary scenario quantum gravity phenomena could be observed in anisotropy 
experiments of the microwave background as well as in galaxy clustering data. 
In particular, according to the latest release of {\em Planck} data the inflationary scale is
significantly lower than the Planck scale, with $k\sim 10^{14}-10^{15}$ GeV for a pivot
scale $k_\star = 0.05$ Mpc$^{-1}$.  An effective action for inflation can then be obtained 
by linearizing the flow around the NGFP and identifying the cutoff with the 
field strength. Starting from a scale-dependent Lagrangian including the (projection of) the currently known relevant coupling constants,
\be
{\mathcal L}_k = \frac{1}{16\pi G_k} \left(R - 2 \Lambda_k \right) - \beta_k \, R^2 \, , 
\ee
and implementing a Type III cutoff identification, a detailed calculation
shows that an effective action valid around the inflationary scale 
may be given by \cite{bo12,bopa15}
\be
\label{eff}
S=\frac{1}{2\kappa^2}\int d^4 x \sqrt{-g} \;\left [ R + \alpha R^{2-\frac{\theta_3}{2}}+ \frac{R^2}{6 m^2} -\Ls \right ] \, . 
\ee
Here $\kappa^2 = 8 \pi G_0$, the scalaron mass $m$ and $\Ls$ encode the details of the RG trajectory in the Einstein-Hilbert sector, and $\theta_3$ is the critical exponent of the $R^2$-operator. The relevant $R^2$-coupling is encoded in $\alpha$.  For $\alpha=0$ eq.\ \eqref{eff} coincides with standard Starobinsky inflation which is favored by the {\em Planck} 2015 data. 
\begin{figure}[t]
	\centering
	\includegraphics[width=8cm]{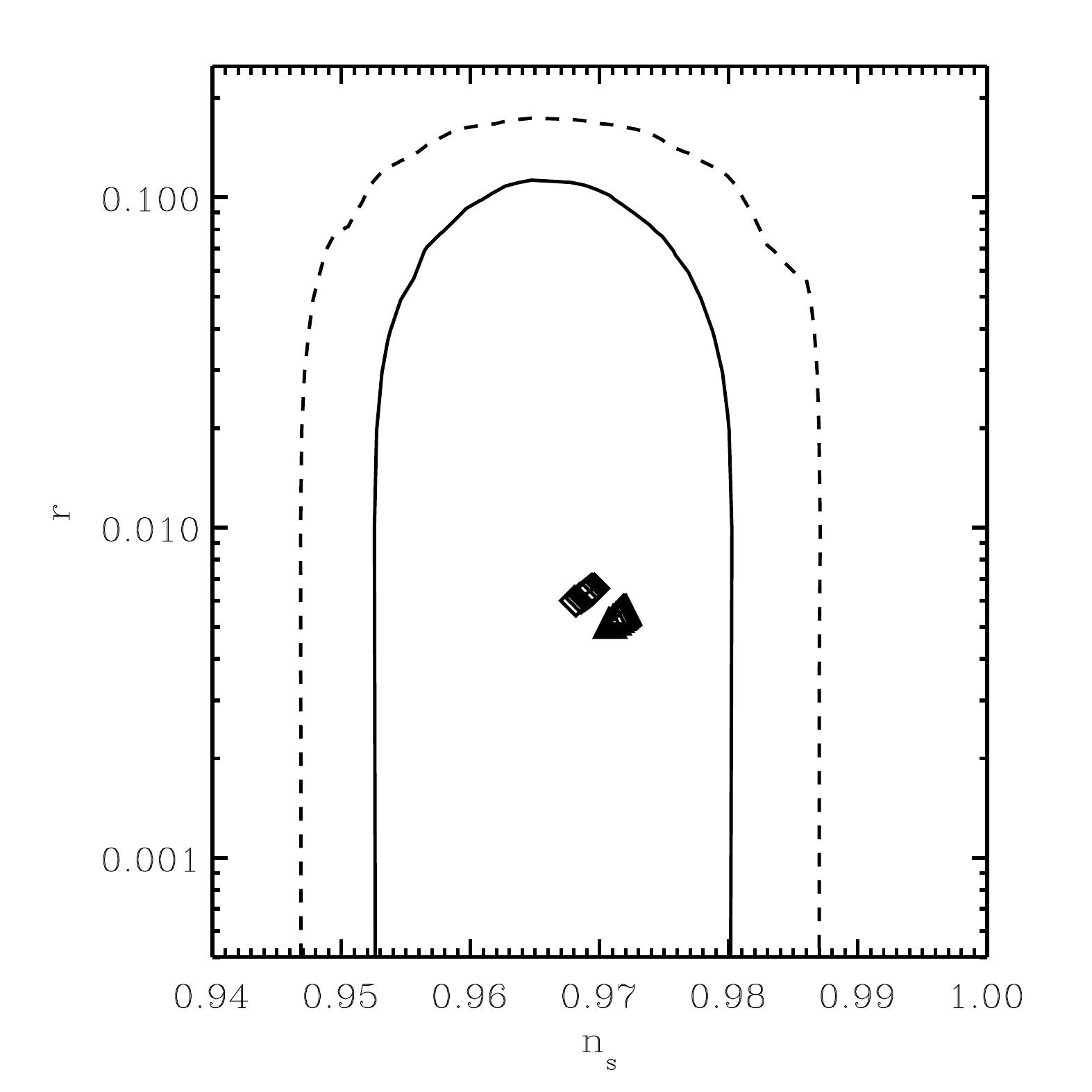}
	\caption{\label{planck}
		Theoretical predictions in the $r$-$n_s$ plane for different values of $\alpha$ for the {\em Planck} collaboration
		2015 data release for the TT correlation assuming $\Lambda$CDM + $r$.
		Triangles are for $N=55$ and squares for $N=60$ e-folds. Solid and dashed lines are the $1\sigma$ and $2\sigma$ confidence levels, respectively 
		(adapted from \cite{bopa15}).
		\label{pla}} 
\end{figure}

It is then possible to constrain the value of $\alpha$ in the slow-roll approximation.  Mapping \eq{eff} to the Einstein frame yields 
\be
\label{effe}	
S=\int d^4 x \sqrt{-g_E} \left [ \frac{1}{2\kappa^2} R_E -\frac{1}{2} g_E^{\mu\nu} \partial_\mu \phi \partial_\nu \phi -V_{\pm}(\phi) \right ] 
\ee
where 
\be\label{effv}
\begin{split}
V_{}(\phi)= & \, \frac{m^2 \mathrm{e}^{-2\sqrt{\frac{2}{3}}\kappa\phi}}{256\kappa^2}\bigg\{192\left(\mathrm{e}^{\sqrt{\frac{2}{3}}\kappa\phi}-1\right)^2-3\alpha^4+128\Ls\\[0.15cm]
&-3\alpha^2\Big(\alpha^2+16\,\mathrm{e}^{\sqrt{\frac{2}{3}}\kappa\phi}-16\Big)- 6\alpha^3\sqrt{\alpha^2+16\,\mathrm{e}^{\sqrt{\frac{2}{3}}\kappa\phi}-16}\\[0.15cm]
&-\sqrt{32}\alpha\Big[\left(\alpha^2+8\mathrm{e}^{\sqrt{\frac{2}{3}}\kappa\phi}-8\right)+\alpha\sqrt{\alpha^2+16\mathrm{e}^{\sqrt{\frac{2}{3}}\kappa\phi}-16}\Big]^{\frac{3}{2}}\bigg\} \,  
\end{split}
\ee
and $g^E_{\mu\nu}=\varphi g_{\mu\nu}$ with $\varphi=e^{\sqrt{2/3}}\kappa\phi$.
In \eq{effv} $\alpha$ and $\Ls$ are in units of the scalaron mass $m$ by means of the rescaling $\alpha \rightarrow \alpha / 3 \sqrt{3} m$
and $\Ls \rightarrow \Ls m^2$, so that both $\alpha$ and $\Ls$ are dimensionless numbers. It is thus possible to constrain the
value of $\alpha$ in the slow-roll approximation so that for $\alpha\in [1,3]$ and $N=50$ e-folds, 
the spectral index $n_s\in (0.965,0.967)$ and the tensor-to-scalar ratio $r\in(0.069,0.0076)$, see Fig.\ \ref{pla}. 
These values significantly larger than the Starobinsky value but still in agreement with observations \cite{Ade:2015lrj}. It is hoped that future CMB anisotropy experiments like CORE \cite{Finelli:2016cyd}, LiteBIRD \cite{Matsumura:2013aja}, or PIXIE \cite{Kogut:2011xw}
 could discriminate among these models.

%------------------------------------------------------------------------
\subsection{Dilaton-gravity models}
\label{sect.4b}
%------------------------------------------------------------------------
The asymptotic safety mechanism, in the case of pure gravity realized through the NGFP \eqref{NGFP}, is also operative in gravity-matter systems \cite{Percacci:2002ie,Dona:2013qba,Meibohm:2015twa}. The occurrence of non-Gaussian gravity-matter fixed points then motivates studying the imprints of Asymptotic Safety also in dilaton-gravity (DG) models. A typical ansatz for the (Euclidean) effective average action reads
\be\label{dilatongrav}
\Gamma_k^{\rm DG} = \int d^4x \sqrt{g} \, \left[ \tfrac{1}{2} F_k(\chi^2) R - \tfrac{1}{2} K_k(\chi^2) \p_\mu \chi \, \p^\mu \chi - V_k(\chi) \right] + \ldots \, , 
\ee 
where the three functions $F,V$ and $K$ depend on the scalar field $\chi$ and the RG scale $k$. Restricting to
\be
F_k(\chi^2) = \frac{1}{16\pi G_k} + \xi_k \, \chi^2 \, , \quad
V_k(\chi^2) = v_k + \tfrac{1}{2} m_k^2 \, \chi^2 + \tfrac{1}{4} \, \sigma_k \, \chi^4 \, , \quad 
K_k = 1 \, ,  
\ee
this class of models also comprises the actions discussed in the context of asymptotically safe Higgs inflation \cite{Xianyu:2014eba,Saltas:2015vsc}.
Substituting the ansatz \eqref{dilatongrav} into the FRGE \eqref{FRGE} yields a system of coupled partial differential equations determining the scale-dependence of $F,V$ and $K$. These equations may be used to integrate down the fixed point potentials to the scale of inflation. Converting to the Einstein frame the predictions for cosmological observables may then be constructed in the standard way.

An alternative approach taken in \cite{Henz:2016aoh} aims at constructing so-called scaling solutions where, by definition, the dimensionless counterparts of $F,V$ and $K$ are independent of the RG scale $k$. Such solutions can be constructed through a combination of analytic and numerical methods. Converting back to the Einstein frame, it is found that the scalar potential is, firstly, independent of the RG scale $k$ and, secondly, possesses a maximum for small values of the scalar field. While these solutions do not (yet) give rise to realistic cosmological models, they serve as a prototype for connecting solutions of the FRGE to cosmology without invoking a cutoff identification. 

%------------------------------------------------------------------------
\section{Summary and outlook}
\label{sect.5}
%------------------------------------------------------------------------
Asymptotic Safety provides an attractive mechanism for constructing a theory of gravity extending to length scales below the Planck scale. The non-Gaussian fixed point controlling the short-distance behavior of the theory leads to a distinct scale-dependence of the gravitational couplings at (trans-)Planckian energies. At the level of cosmological model building, this scale-dependence may be taken into account via a RG improvement either at the level of the equations of motion or the effective average action. The resulting models may naturally give rise to a quantum gravity driven inflationary phase in the very early universe, generate a significant amount of entropy through an energy transfer from the gravitational to the matter sector, and should possess an (almost) scale-free fluctuation spectrum \cite{Bonanno:2001xi,Reuter:2005kb,Bonanno:2007wg,Koch:2010nn}. Moreover, effective actions arising from the RG improvement permit a direct comparison to cosmological observations as, e.g.\ \emph{Planck} data, potentially constraining the relevant couplings of the theory. Generically, the RG improvements studied so far do not resolve the initial Big Bang singularity. There are indications, however, that in some cases the singularity is replaced by a bounce \cite{Kofinas:2016lcz}, though. While a derivation of fluctuation spectra based on a first principle computation in Asymptotic Safety is still missing, the construction of consistent RG flows in a Friedmann-Robertson-Walker background recently completed in \cite{Biemans:2016rvp} constitutes an important first step in this direction. 

\bigskip
\noindent
\emph{Acknowledgments} The research of F.~S.\
is supported by the Netherlands Organisation for Scientific
Research (NWO) within the Foundation for Fundamental Research on Matter (FOM) grants 13PR3137 and 13VP12.
%------------------------------------------------------------------------
\section*{References}
%------------------------------------------------------------------------


\begin{thebibliography}{10}
	\expandafter\ifx\csname url\endcsname\relax
	\def\url#1{\texttt{#1}}\fi
	\expandafter\ifx\csname urlprefix\endcsname\relax\def\urlprefix{URL }\fi
	\expandafter\ifx\csname href\endcsname\relax
	\def\href#1#2{#2} \def\path#1{#1}\fi
	
	\bibitem{Weinberg:2009bg}
	S.~Weinberg,
	% {Effective Field Theory, Past and Future}, 
	PoS CD09 (2009) 001,
	\newblock \href {http://arxiv.org/abs/0908.1964} {\path{arXiv:0908.1964}}.
	
	\bibitem{Reuter:1996cp}
	M.~Reuter, 
	%{Nonperturbative evolution equation for quantum gravity}, 
	Phys.\ Rev.\ D57 (1998) 971, % --985.
	\newblock \href {http://arxiv.org/abs/hep-th/9605030}
	{\path{arXiv:hep-th/9605030}}. %, \href
%	{http://dx.doi.org/10.1103/PhysRevD.57.971}
%	{\path{doi:10.1103/PhysRevD.57.971}}.
	
	\bibitem{Niedermaier:2006wt}
	M.~Niedermaier, M.~Reuter, 
	%{The Asymptotic Safety Scenario in Quantum Gravity},
	Living Rev.\ Rel.\ 9 (2006) 5.
%	\newblock \href {http://dx.doi.org/10.12942/lrr-2006-5}
%	{\path{doi:10.12942/lrr-2006-5}}.
	
	\bibitem{Codello:2008vh}
	A.~Codello, R.~Percacci, C.~Rahmede,
	% {Investigating the Ultraviolet Properties
	%	of Gravity with a Wilsonian Renormalization Group Equation},
	 Annals Phys.\ 324
	(2009) 414, %--469.
	\newblock \href {http://arxiv.org/abs/0805.2909} {\path{arXiv:0805.2909}}.
%	\href {http://dx.doi.org/10.1016/j.aop.2008.08.008}
%	{\path{doi:10.1016/j.aop.2008.08.008}}.
	
	\bibitem{Litim:2011cp}
	D.~F.~Litim, 
	%{Renormalisation group and the Planck scale}, 
	Phil.\ Trans.\ Roy.\ Soc.\ Lond.\ A369 (2011) 2759,
	\newblock \href {http://arxiv.org/abs/1102.4624} {\path{arXiv:1102.4624}}.
%	\href {http://dx.doi.org/10.1098/rsta.2011.0103}
%	{\path{doi:10.1098/rsta.2011.0103}}.
	
	\bibitem{Percacci:2011fr}
	R.~Percacci,
%	\href{http://inspirehep.net/record/943400/files/arXiv:1110.6389.pdf}{{A Short
%			introduction to asymptotic safety}}, in: {Time and Matter}, 2011, pp.
%	123--142.
	\newblock \href {http://arxiv.org/abs/1110.6389} {\path{arXiv:1110.6389}}.
%	\newline\urlprefix\url{http://inspirehep.net/record/943400/files/arXiv:1110.6389.pdf}
	
	\bibitem{Reuter:2012id}
	M.~Reuter, F.~Saueressig,
	% {Quantum Einstein Gravity},
	New J.\ Phys.\ 14 (2012) 055022,
	\newblock \href {http://arxiv.org/abs/1202.2274} {\path{arXiv:1202.2274}}.
%	\href {http://dx.doi.org/10.1088/1367-2630/14/5/055022}
%	{\path{doi:10.1088/1367-2630/14/5/055022}}.
	
	\bibitem{Reuter:2012xf}
	M.~Reuter, F.~Saueressig,
	% {Asymptotic Safety, Fractals, and Cosmology}, 
	Lect.\ 	Notes Phys.\ 863 (2013) 185,
	\newblock \href {http://arxiv.org/abs/1205.5431} {\path{arXiv:1205.5431}}.
%	\href {http://dx.doi.org/10.1007/978-3-642-33036-0_8}
%	{\path{doi:10.1007/978-3-642-33036-0_8}}.
	
	\bibitem{Bonanno:2001xi}
	A.~Bonanno, M.~Reuter,
	% {Cosmology of the Planck era from a renormalization
	%	group for quantum gravity}, 
	Phys.\ Rev.\ D65 (2002) 043508,
	\newblock \href {http://arxiv.org/abs/hep-th/0106133}
	{\path{arXiv:hep-th/0106133}}.
	%, \href
	%{http://dx.doi.org/10.1103/PhysRevD.65.043508}
	%{\path{doi:10.1103/PhysRevD.65.043508}}.
	
	\bibitem{Bonanno:2001hi}
	A.~Bonanno, M.~Reuter, 
	%{Cosmology with selfadjusting vacuum energy density from
	%	a renormalization group fixed point},
	 Phys.\ Lett.\ B527 (2002) 9,
	\newblock \href {http://arxiv.org/abs/astro-ph/0106468}
	{\path{arXiv:astro-ph/0106468}}.
	%, \href
	%{http://dx.doi.org/10.1016/S0370-2693(01)01522-2}
	%{\path{doi:10.1016/S0370-2693(01)01522-2}}.
	
	\bibitem{Bentivegna:2003rr}
	E.~Bentivegna, A.~Bonanno, M.~Reuter,
	% {Confronting the IR fixed point cosmology
	%	with high redshift supernova data},
	 JCAP 01 (2004) 001,
	\newblock \href {http://arxiv.org/abs/astro-ph/0303150}
	{\path{arXiv:astro-ph/0303150}}. %, \href
%	{http://dx.doi.org/10.1088/1475-7516/2004/01/001}
%	{\path{doi:10.1088/1475-7516/2004/01/001}}.
	
	\bibitem{Reuter:2005kb}
	M.~Reuter, F.~Saueressig, %{From big bang to asymptotic de Sitter: Complete
	%	cosmologies in a quantum gravity framework},
	 JCAP 09 (2005) 012,
	\newblock \href {http://arxiv.org/abs/hep-th/0507167}
	{\path{arXiv:hep-th/0507167}}. %, \href
%	{http://dx.doi.org/10.1088/1475-7516/2005/09/012}
%	{\path{doi:10.1088/1475-7516/2005/09/012}}.
	
	\bibitem{Bonanno:2007wg}
	A.~Bonanno, M.~Reuter, 
	%{Entropy signature of the running cosmological
	%	constant}, 
	JCAP 08 (2007) 024,
	\newblock \href {http://arxiv.org/abs/0706.0174} {\path{arXiv:0706.0174}}.
%	\href {http://dx.doi.org/10.1088/1475-7516/2007/08/024}
%	{\path{doi:10.1088/1475-7516/2007/08/024}}.
	
	\bibitem{Weinberg:2009wa}
	S.~Weinberg, 
	%{Asymptotically Safe Inflation},
	 Phys.\ Rev.\ D81 (2010) 083535,
	\newblock \href {http://arxiv.org/abs/0911.3165} {\path{arXiv:0911.3165}}.
%	\href {http://dx.doi.org/10.1103/PhysRevD.81.083535}
%	{\path{doi:10.1103/PhysRevD.81.083535}}.
	
	\bibitem{Bonanno:2009nj}
	A.~Bonanno, 
	%{Astrophysical implications of the Asymptotic Safety Scenario in
	%	Quantum Gravity}, 
	PoS CLAQG08 (2011) 008,
	\newblock \href {http://arxiv.org/abs/0911.2727} {\path{arXiv:0911.2727}}.
	
	\bibitem{Bonanno:2010mk}
	A.~Bonanno, M.~Reuter, 
	%{Entropy Production during Asymptotically Safe
	%	Inflation}, 
	Entropy 13 (2011) 274,
	\newblock \href {http://arxiv.org/abs/1011.2794} {\path{arXiv:1011.2794}}.
%	\href {http://dx.doi.org/10.3390/e13010274} {\path{doi:10.3390/e13010274}}.
	
	\bibitem{Koch:2010nn}
	B.~Koch, I.~Ramirez,
	% {Exact renormalization group with optimal scale and its
	%	application to cosmology},
	 Class.\ Quant.\ Grav.\ 28 (2011) 055008,
	\newblock \href {http://arxiv.org/abs/1010.2799} {\path{arXiv:1010.2799}}.
%	\href {http://dx.doi.org/10.1088/0264-9381/28/5/055008}
%	{\path{doi:10.1088/0264-9381/28/5/055008}}.
	
	\bibitem{Casadio:2010fw}
	R.~Casadio, S.~D.~H. Hsu, B.~Mirza, 
	%{Asymptotic Safety, Singularities, and
	%	Gravitational Collapse}, 
	Phys.\ Lett.\ B695 (2011) 317, %--319.
	\newblock \href {http://arxiv.org/abs/1008.2768} {\path{arXiv:1008.2768}}.
%	\href {http://dx.doi.org/10.1016/j.physletb.2010.10.060}
%	{\path{doi:10.1016/j.physletb.2010.10.060}}.
	
	\bibitem{Contillo:2010ju}
	A.~Contillo, 
	%{Evolution of cosmological perturbations in an RG-driven
	%	inflationary scenario},
	 Phys.\ Rev.\ D83 (2011) 085016,
	\newblock \href {http://arxiv.org/abs/1011.4618} {\path{arXiv:1011.4618}}.
%	\href {http://dx.doi.org/10.1103/PhysRevD.83.085016}
%	{\path{doi:10.1103/PhysRevD.83.085016}}.
	
	\bibitem{Bonanno:2010bt}
	A.~Bonanno, A.~Contillo, R.~Percacci,
	% {Inflationary solutions in asymptotically
	%	safe f(R) theories}, 
	Class.\ Quant.\ Grav.\ 28 (2011) 145026,
	\newblock \href {http://arxiv.org/abs/1006.0192} {\path{arXiv:1006.0192}}.
%	\href {http://dx.doi.org/10.1088/0264-9381/28/14/145026}
%	{\path{doi:10.1088/0264-9381/28/14/145026}}.
	
	\bibitem{Frolov:2011ys}
	A.~V.~Frolov, J.-Q.~Guo, 
	%{Small Cosmological Constant from Running
	%	Gravitational Coupling}
	\href {http://arxiv.org/abs/1101.4995}
	{\path{arXiv:1101.4995}}.
	
	\bibitem{Hindmarsh:2011hx}
	M.~Hindmarsh, D.~Litim, C.~Rahmede,
	% {Asymptotically Safe Cosmology}, 
	JCAP 07 (2011) 019,
	\newblock \href {http://arxiv.org/abs/1101.5401} {\path{arXiv:1101.5401}}.
%	\href {http://dx.doi.org/10.1088/1475-7516/2011/07/019}
%	{\path{doi:10.1088/1475-7516/2011/07/019}}.
	
	\bibitem{Bonanno:2011yx}
	A.~Bonanno, S.~Carloni, 
	%{Dynamical System Analysis of Cosmologies with Running
	%	Cosmological Constant from Quantum Einstein Gravity}, 
	New J.\ Phys.\ 14 (2012) 025008,
	\newblock \href {http://arxiv.org/abs/1112.4613} {\path{arXiv:1112.4613}}.
%	\href {http://dx.doi.org/10.1088/1367-2630/14/2/025008}
%	{\path{doi:10.1088/1367-2630/14/2/025008}}.
	
	\bibitem{Ahn:2011qt}
	C.~Ahn, C.~Kim, E.~V. Linder, 
	%{From Asymptotic Safety to Dark Energy}, 
	Phys.\ Lett.\ B704 (2011) 10,
	\newblock \href {http://arxiv.org/abs/1106.1435} {\path{arXiv:1106.1435}}.
%	\href {http://dx.doi.org/10.1016/j.physletb.2011.08.075}
%	{\path{doi:10.1016/j.physletb.2011.08.075}}.
	
	\bibitem{Cai:2011kd}
	Y.-F.~Cai, D.~A.~Easson,
	% {Asymptotically safe gravity as a scalar-tensor theory
	%	and its cosmological implications},
	 Phys.\ Rev.\ D84 (2011) 103502,
	\newblock \href {http://arxiv.org/abs/1107.5815} {\path{arXiv:1107.5815}}.
%	\href {http://dx.doi.org/10.1103/PhysRevD.84.103502}
%	{\path{doi:10.1103/PhysRevD.84.103502}}.
	
	\bibitem{Contillo:2011ag}
	A.~Contillo, M.~Hindmarsh, C.~Rahmede, 
	%{Renormalisation group improvement of
	%	scalar field inflation},
	 Phys. Rev. D85 (2012) 043501,
	\newblock \href {http://arxiv.org/abs/1108.0422} {\path{arXiv:1108.0422}}.
%	\href {http://dx.doi.org/10.1103/PhysRevD.85.043501}
%	{\path{doi:10.1103/PhysRevD.85.043501}}.
	
	\bibitem{Cai:2012qi}
	Y.-F.~Cai, D.~A.~Easson,
	% {Higgs Boson in RG running Inflationary Cosmology},
	Int.\ J.\ Mod.\ Phys.\ D21 (2013) 1250094,
	\newblock \href {http://arxiv.org/abs/1202.1285} {\path{arXiv:1202.1285}}.
%	\href {http://dx.doi.org/10.1142/S0218271812500940}
%	{\path{doi:10.1142/S0218271812500940}}.
	
	\bibitem{Bonanno:2012jy}
	A.~Bonanno, 
	%{An effective action for asymptotically safe gravity},
	Phys.\ Rev.\ D85 (2012) 081503,
	\newblock \href {http://arxiv.org/abs/1203.1962} {\path{arXiv:1203.1962}}.
%	\href {http://dx.doi.org/10.1103/PhysRevD.85.081503}
%	{\path{doi:10.1103/PhysRevD.85.081503}}.
	
	\bibitem{Hindmarsh:2012rc}
	M.~Hindmarsh, I.~D. Saltas, 
	%{f(R) Gravity from the renormalisation group},
	Phys.\ Rev.\ D86 (2012) 064029,
	\newblock \href {http://arxiv.org/abs/1203.3957} {\path{arXiv:1203.3957}}.
%	\href {http://dx.doi.org/10.1103/PhysRevD.86.064029}
%	{\path{doi:10.1103/PhysRevD.86.064029}}.
	
	\bibitem{Fang:2012ca}
	C.~Fang, Q.-G. Huang, 
%	{The trouble with asymptotically safe inflation}, 
	Eur.\ 	Phys.\ J.\ C73 (2013) 2401,
	\newblock \href {http://arxiv.org/abs/1210.7596} {\path{arXiv:1210.7596}}.
%	\href {http://dx.doi.org/10.1140/epjc/s10052-013-2401-2}
%	{\path{doi:10.1140/epjc/s10052-013-2401-2}}.
	
	\bibitem{Bonanno:2013dja}
	A.~Bonanno, M.~Reuter, 
	%{Modulated Ground State of Gravity Theories with
	%	Stabilized Conformal Factor}, 
	Phys.\ Rev.\ D87 (2013) 084019,
	\newblock \href {http://arxiv.org/abs/1302.2928} {\path{arXiv:1302.2928}}.
%	\href {http://dx.doi.org/10.1103/PhysRevD.87.084019}
%	{\path{doi:10.1103/PhysRevD.87.084019}}.
	
	\bibitem{Copeland:2013vva}
	E.~J. Copeland, C.~Rahmede, I.~D. Saltas,
	% {Asymptotically Safe Starobinsky 	Inflation}, 
	Phys.\ Rev.\ D91 (2015) 103530,
	\newblock \href {http://arxiv.org/abs/1311.0881} {\path{arXiv:1311.0881}}.
%	\href {http://dx.doi.org/10.1103/PhysRevD.91.103530}
%	{\path{doi:10.1103/PhysRevD.91.103530}}.
	
	\bibitem{Kaya:2013bga}
	A.~Kaya, 
	%{Exact renormalization group flow in an expanding Universe and
	%	screening of the cosmological constant}, 
	Phys.\ Rev.\ D87 (2013) 123501,
	\newblock \href {http://arxiv.org/abs/1303.5459} {\path{arXiv:1303.5459}}.
%	\href {http://dx.doi.org/10.1103/PhysRevD.87.123501}
%	{\path{doi:10.1103/PhysRevD.87.123501}}.
	
	\bibitem{Becker:2014jua}
	D.~Becker, M.~Reuter, 
	%{Propagating gravitons vs. 'dark matter` in
	%	asymptotically safe quantum gravity}, 
	JHEP 12 (2014) 025,
	\newblock \href {http://arxiv.org/abs/1407.5848} {\path{arXiv:1407.5848}}.
	%\href {http://dx.doi.org/10.1007/JHEP12(2014)025}
%	{\path{doi:10.1007/JHEP12(2014)025}}.
	
	\bibitem{Xianyu:2014eba}
	Z.-Z. Xianyu, H.-J. He,
	% {Asymptotically Safe Higgs Inflation}, 
	JCAP 10 (2014) 083,
	\newblock \href {http://arxiv.org/abs/1407.6993} {\path{arXiv:1407.6993}}.
%	\href {http://dx.doi.org/10.1088/1475-7516/2014/10/083}
%	{\path{doi:10.1088/1475-7516/2014/10/083}}.
	
	\bibitem{Saltas:2015vsc}
	I.~D. Saltas, 
	%{Higgs inflation and quantum gravity: An exact renormalisation
	%	group approach}, 
	JCAP 02 (2016) 048,
	\newblock \href {http://arxiv.org/abs/1512.06134} {\path{arXiv:1512.06134}}.
%	\href {http://dx.doi.org/10.1088/1475-7516/2016/02/048}
%	{\path{doi:10.1088/1475-7516/2016/02/048}}.
	
	\bibitem{Nielsen:2015una}
	N.~G. Nielsen, F.~Sannino, O.~Svendsen, 
	%{Inflation from Asymptotically Safe Theories}, 
	Phys.\ Rev.\ D91 (2015) 103521,
	\newblock \href {http://arxiv.org/abs/1503.00702} {\path{arXiv:1503.00702}}.
%	\href {http://dx.doi.org/10.1103/PhysRevD.91.103521}
%	{\path{doi:10.1103/PhysRevD.91.103521}}.
	
	\bibitem{Bonanno:2015fga}
	A.~Bonanno, A.~Platania, 
	%{Asymptotically safe inflation from quadratic gravity}, 
	Phys.\ Lett.\ B750 (2015) 638, %--642.
	\newblock \href {http://arxiv.org/abs/1507.03375} {\path{arXiv:1507.03375}}.
%	\href {http://dx.doi.org/10.1016/j.physletb.2015.10.005}
%	{\path{doi:10.1016/j.physletb.2015.10.005}}.
	
	\bibitem{Benedetti:2010nr}
	D.~Benedetti, K.~Groh, P.~F. Machado, F.~Saueressig,
	% {The Universal RG Machine}, 
	JHEP 06 (2011) 079,
	\newblock \href {http://arxiv.org/abs/1012.3081} {\path{arXiv:1012.3081}}.
%	\href {http://dx.doi.org/10.1007/JHEP06(2011)079}
%	{\path{doi:10.1007/JHEP06(2011)079}}.
	
	%\cite{Litim:2001up}
	\bibitem{Litim:2001up}
	D.~F.~Litim,
	%``Optimized renormalization group flows,''
	Phys.\ Rev.\ D64 (2001) 105007, 
	\newblock \href {http://arxiv.org/abs/hep-th/0103195} {\path{hep-th/0103195}}.
	%%CITATION = doi:10.1103/PhysRevD.64.105007;%%
	%432 citations counted in INSPIRE as of 30 Oct 2016
	
	%\cite{Reuter:2001ag}
	\bibitem{Reuter:2001ag}
	M.~Reuter and F.~Saueressig,
	%``Renormalization group flow of quantum gravity in the Einstein-Hilbert truncation,''
	Phys.\ Rev.\ D {\bf 65} (2002) 065016,
		\newblock \href {http://arxiv.org/abs/hep-th/0110054} {\path{hep-th/0110054}}.
	%%CITATION = doi:10.1103/PhysRevD.65.065016;%%
	%251 citations counted in INSPIRE as of 30 Oct 2016
	
	\bibitem{Reuter:2004nx}
	M.~Reuter, H.~Weyer,
	% {Quantum gravity at astrophysical distances?},
	 JCAP 12
	(2004) 001,
	\newblock \href {http://arxiv.org/abs/hep-th/0410119}
	{\path{arXiv:hep-th/0410119}}.
	%, \href
	%{http://dx.doi.org/10.1088/1475-7516/2004/12/001}
	%{\path{doi:10.1088/1475-7516/2004/12/001}}.
	
	%\cite{Lauscher:2002sq}
	\bibitem{Lauscher:2002sq}
	O.~Lauscher and M.~Reuter,
	%``Flow equation of quantum Einstein gravity in a higher derivative truncation,''
	Phys.\ Rev.\ D {\bf 66} (2002) 025026,
%	doi:10.1103/PhysRevD.66.025026
	\newblock \href {http://arxiv.org/abs/hep-th/0205062}
	{\path{arXiv:hep-th/0205062}}.
	%%CITATION = doi:10.1103/PhysRevD.66.025026;%%
	%229 citations counted in INSPIRE as of 30 Oct 2016
	
	\bibitem{Benedetti:2009rx}
	D.~Benedetti, P.~F. Machado, F.~Saueressig, 
	%{Asymptotic safety in
	%	higher-derivative gravity}, 
	Mod.\ Phys.\ Lett.\ A24 (2009) 2233, %--2241.
	\newblock \href {http://arxiv.org/abs/0901.2984} {\path{arXiv:0901.2984}},
%	\href {http://dx.doi.org/10.1142/S0217732309031521}
%	{\path{doi:10.1142/S0217732309031521}}.
	
	
	
	\bibitem{Rechenberger:2012pm}
	S.~Rechenberger, F.~Saueressig,
	% {The $R^2$ phase-diagram of QEG and its
	%	spectral dimension}, 
	Phys.\ Rev.\ D86 (2012) 024018,
	\newblock \href {http://arxiv.org/abs/1206.0657} {\path{arXiv:1206.0657}},
%	\href {http://dx.doi.org/10.1103/PhysRevD.86.024018}
%	{\path{doi:10.1103/PhysRevD.86.024018}}.
	
	\bibitem{Gies:2016con}
	H.~Gies, B.~Knorr, S.~Lippoldt, F.~Saueressig, 
	%{Gravitational Two-Loop
	%	Counterterm Is Asymptotically Safe}, 
	Phys.\ Rev.\ Lett.\ 116 (2016) 211302,
	\newblock \href {http://arxiv.org/abs/1601.01800} {\path{arXiv:1601.01800}},
	%\href {http://dx.doi.org/10.1103/PhysRevLett.116.211302}
	%{\path{doi:10.1103/PhysRevLett.116.211302}}.
	
	\bibitem{Migdal:1973si}
	A.~B. Migdal, 
	%{Vacuum polarization in strong non-homogeneous fields},
	Nucl.\ 	Phys.\ B52 (1973) 483. %--505.
%	\newblock \href {http://dx.doi.org/10.1016/0550-3213(73)90575-0}
%	{\path{doi:10.1016/0550-3213(73)90575-0}}.
	
	\bibitem{Adler:1982js}
	S.~L.~Adler, 
	%{DYNAMICAL APPLICATIONS OF THE GAUGE INVARIANT EFFECTIVE ACTION	FORMALISM}.
	in S.~M.~Christensen (Ed.), \emph{Quantum Theory Of Gravity}, Adam Hilger Ltd., Bristol (1984) 388.
	
	\bibitem{Dittrich:1985yb}
	W.~Dittrich, M.~Reuter, %{EFFECTIVE LAGRANGIANS IN QUANTUM ELECTRODYNAMICS},
	Lect.\ Notes Phys.\ 220 (1985) 1.
	
	\bibitem{Kofinas:2016lcz}
	G.~Kofinas, V.~Zarikas, 
	%{Asymptotic Safe gravity and non-singular inflationary
	%	Big Bang with vacuum birth}
	\href {http://arxiv.org/abs/1605.02241}
	{\path{arXiv:1605.02241}}.
	
	%\cite{Antoniadis:1996dj}
	\bibitem{Antoniadis:1996dj}
	I.~Antoniadis, P.~O.~Mazur and E.~Mottola,
	%``Conformal invariance and cosmic background radiation,''
	Phys.\ Rev.\ Lett.\  {\bf 79} (1997) 14,
%	doi:10.1103/PhysRevLett.79.14
	\href {https://arxiv.org/pdf/astro-ph/9611208.pdf}
	{\path{astro-ph/9611208}}.
	%%CITATION = doi:10.1103/PhysRevLett.79.14;%%
	%69 citations counted in INSPIRE as of 23 Jan 2017
	
	\bibitem{D'Odorico:2015lhd}
	G.~D'Odorico, F.~Saueressig,
	% {Quantum phase transitions in the
	%	Belinsky-Khalatnikov-Lifshitz universe}, 
	Phys. Rev. D92 (2015) 124068,
	\newblock \href {http://arxiv.org/abs/1511.00247} {\path{arXiv:1511.00247}}.
%	\href {http://dx.doi.org/10.1103/PhysRevD.92.124068}
%	{\path{doi:10.1103/PhysRevD.92.124068}}.
	
	\bibitem{Belinsky:1970ew}
	V.~A. Belinsky, I.~M. Khalatnikov, E.~M. Lifshitz, 
	%{Oscillatory approach to a
	%	singular point in the relativistic cosmology}, 
	Adv.\ Phys.\ 19 (1970) 525. %--573.
%	\newblock \href {http://dx.doi.org/10.1080/00018737000101171}
%	{\path{doi:10.1080/00018737000101171}}.
	
	\bibitem{Belinski:1973zz}
	V.~A. Belinski, I.~M. Khalatnikov, 
	%{Effect of Scalar and Vector Fields on the
	%	Nature of the Cosmological Singularity}, 
	Sov.\ Phys.\ JETP 36 (1973) 591.
	
	\bibitem{Belinsky:1982pk}
	V.~A.~Belinsky, I.~M.~Khalatnikov, E.~M.~Lifshitz,
	% {A General Solution of the
%		Einstein Equations with a Time Singularity}, 
	Adv.\ Phys.\ 31 (1982) 639. %--667.
%	\newblock \href {http://dx.doi.org/10.1080/00018738200101428}
%	{\path{doi:10.1080/00018738200101428}}.
	
	\bibitem{Andersson:2000cv}
	L.~Andersson, A.~D. Rendall, 
	%{Quiescent cosmological singularities}, 
	Commun.\ Math.\ Phys.\ 218 (2001) 479, 
	\newblock \href {http://arxiv.org/abs/gr-qc/0001047}
	{\path{arXiv:gr-qc/0001047}}.
	%, \href {http://dx.doi.org/10.1007/s002200100406}
	%{\path{doi:10.1007/s002200100406}}.
	
	\bibitem{Koslowski:2016hds}
	T.~A. Koslowski, F.~Mercati, D.~Sloan, 
	%{Relationalism Evolves the Universe
	%	Through the Big Bang}
	\href {http://arxiv.org/abs/1607.02460}
	{\path{arXiv:1607.02460}}.
	
	\bibitem{bo12}
	A.~Bonanno, 
	%{An effective action for asymptotically safe gravity}, 
	Phys.\ Rev.\ D85 (2012) 081503,
	\newblock \href {http://arxiv.org/abs/1203.1962} {\path{arXiv:1203.1962}},
%	\href {http://dx.doi.org/10.1103/PhysRevD.85.081503}
%	{\path{doi:10.1103/PhysRevD.85.081503}}.
	
	\bibitem{bopa15}
	A.~Bonanno, A.~Platania,
	% {Asymptotically safe inflation from quadratic gravity}, 
	Phys.\ Lett.\ B750 (2015) 638,
	\newblock \href {http://arxiv.org/abs/1507.03375} {\path{arXiv:1507.03375}},
%	\href {http://dx.doi.org/10.1016/j.physletb.2015.10.005}
%	{\path{doi:10.1016/j.physletb.2015.10.005}}.
	
	\bibitem{Ade:2015lrj}
	P.~A.~R. Ade, et~al., 
	%{Planck 2015 results. XX. Constraints on inflation},
	Astron.\ Astrophys.\ 594 (2016) A20,
	\newblock \href {http://arxiv.org/abs/1502.02114} {\path{arXiv:1502.02114}},
%	\href {http://dx.doi.org/10.1051/0004-6361/201525898}
%	{\path{doi:10.1051/0004-6361/201525898}}.
	
%\cite{Finelli:2016cyd}
\bibitem{Finelli:2016cyd}
F.~Finelli {\it et al.} [CORE Collaboration],
%``Exploring Cosmic Origins with CORE: Inflation,''
\newblock \href {https://arxiv.org/pdf/1612.08270.pdf} {\path{arXiv:1612.08270}}.
%%CITATION = ARXIV:1612.08270;%%

%\cite{Matsumura:2013aja}
\bibitem{Matsumura:2013aja}
T.~Matsumura {\it et al.},
%``Mission design of LiteBIRD,''
J.\ Low.\ Temp.\ Phys.\  {\bf 176} (2014) 733,
\newblock \href {https://arxiv.org/pdf/1311.2847.pdf} {\path{arXiv:1311.2847}}.
%doi:10.1007/s10909-013-0996-1
%[arXiv:1311.2847 [astro-ph.IM]].
%%CITATION = doi:10.1007/s10909-013-0996-1;%%
%82 citations counted in INSPIRE as of 23 Jan 2017

%\cite{Kogut:2011xw}
\bibitem{Kogut:2011xw}
A.~Kogut {\it et al.},
%``The Primordial Inflation Explorer (PIXIE): A Nulling Polarimeter for Cosmic Microwave Background Observations,''
JCAP {\bf 1107} (2011) 025,
%doi:10.1088/1475-7516/2011/07/025
\newblock \href {https://arxiv.org/pdf/1105.2044.pdf} {\path{arXiv:1105.2044}}.
%[arXiv:1105.2044 [astro-ph.CO]].
%%CITATION = doi:10.1088/1475-7516/2011/07/025;%%
%237 citations counted in INSPIRE as of 23 Jan 2017	
	
	%\cite{Percacci:2002ie}
	\bibitem{Percacci:2002ie}
	R.~Percacci and D.~Perini,
	%``Constraints on matter from asymptotic safety,''
	Phys.\ Rev.\ D {\bf 67} (2003) 081503,
		\newblock \href {http://arxiv.org/abs/hep-th/0207033}
		{\path{arXiv:hep-th/0207033}}.
	%%CITATION = doi:10.1103/PhysRevD.67.081503;%%
	%154 citations counted in INSPIRE as of 30 Oct 2016
	
	\bibitem{Dona:2013qba}
	P.~Dona, A.~Eichhorn, R.~Percacci, 
	%{Matter matters in asymptotically safe quantum gravity}, 
	Phys.\ Rev.\ D89 (2014) 084035,
	\newblock \href {http://arxiv.org/abs/1311.2898} {\path{arXiv:1311.2898}},
%	\href {http://dx.doi.org/10.1103/PhysRevD.89.084035}
%	{\path{doi:10.1103/PhysRevD.89.084035}}.
	
	\bibitem{Meibohm:2015twa}
	J.~Meibohm, J.~M. Pawlowski, M.~Reichert, 
	%{Asymptotic safety of gravity-matter systems},
	 Phys.\ Rev.\ D93 (2016) 084035,
	\newblock \href {http://arxiv.org/abs/1510.07018} {\path{arXiv:1510.07018}},
%	\href {http://dx.doi.org/10.1103/PhysRevD.93.084035}
%	{\path{doi:10.1103/PhysRevD.93.084035}}.
	
	\bibitem{Henz:2016aoh}
	T.~Henz, J.~M. Pawlowski, C.~Wetterich,
	% {Scaling solutions for Dilaton Quantum Gravity}
	\href {http://arxiv.org/abs/1605.01858} {\path{arXiv:1605.01858}}.
	
	\bibitem{Biemans:2016rvp}
	J.~Biemans, A.~Platania, F.~Saueressig,
	% {Quantum gravity on foliated spacetime
	%	- asymptotically safe and sound}
	\href{http://arxiv.org/abs/1609.04813}{\path{arXiv:1609.04813}}.


	
\end{thebibliography}
\end{document}